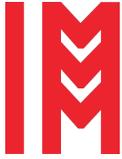

# Which Trading Agent is Best? Using a Threaded Parallel Simulation of a Financial Market Changes the Pecking-Order

Michael Rollins and Dave Cliff*

Department of Computer Science, The University of Bristol, Woodland Road, Bristol, BS8 1UB, UK

*Corresponding author. Email address: csdtc@bristol.ac.uk

**Abstract**

This paper presents novel results, generated from a new simulation model of a contemporary financial market, that cast serious doubt on the previously widely accepted view of the relative performance of various well-known public-domain automated-trading algorithms. Put simply, we show here that if you use a more realistic market simulator, then trading algorithms previously thought to be the best-performing are shown to be not as good as people think they are, and some algorithms previously thought to be poor performers can be seen to do surprisingly well. Automated trading is now entirely commonplace in most of the world's major financial markets: adaptive algorithmic trading systems operate largely autonomously, interacting with other traders (either other automated systems, or humans) via an electronic exchange platform. Various public-domain trading algorithms have been proposed over the past 25 years in a kind of arms-race, where each new trading algorithm was compared to the previous best, thereby establishing a "pecking order", i.e. a partially-ordered dominance hierarchy from best to worst of the various trading algorithms. Many of these algorithms were developed, tested, and evaluated using simple minimal simulations of financial markets that only very weakly approximated the fact that real markets involve many different trading systems operating asynchronously and in parallel. In this paper we use BSE, a long-established public-domain market simulator, to run a set of experiments generating benchmark results from several well-known trading algorithms. BSE incorporates a very simple time-sliced approach to simulating parallelism, which has obvious known weaknesses. We then alter and extend BSE to make it *threaded*, so that different trader algorithms operate asynchronously and in parallel: we call this simulator *Threaded-BSE* (TBSE). We then re-run the trader experiments on TBSE and compare the TBSE results to our earlier benchmark results from BSE. Our comparison shows that the dominance hierarchy in our more realistic experiments is different from the one given by the original simple simulator. We conclude that simulated parallelism matters a lot, and that earlier results from simple simulations comparing different trader algorithms are no longer to be entirely trusted.

## 1. Introduction

This paper presents new evidence that key results previously published in the literature on automated trading systems may be artefacts of the simplistic modelling and simulation methods used to generate them. In particular, we have taken a long-established public-domain simulator of a contemporary electronic financial market, called BSE, and used it to generate a set of benchmark results from four very widely-used public-domain automated-trading algorithms, known as ZIC, ZIP, GDX, and AA (we explain these in more detail, later). These initial BSE benchmark results confirm the previously-published dominance-hierarchy, or "pecking order", where AA beats GDX, GDX beats ZIP, and ZIP beats ZIC, which we write in abbreviated form as AA>GDX>ZIP>ZIC. The BSE simulator is deliberately simple in its modelling of time, and uses only a single thread, a single computer process, which it divides among the various simulated automated traders in such a way that the response-time, the time for an automated trader to compute an answer, is not simulated and hence very complicated trading systems that would take a lot of "thinking time" to compute a response to some change in the market are treated as if they take exactly the same time as the very fastest of trading algorithms. This is clearly unrealistic, yet BSE's approach to simulating time is not uncommon.

We then constructed a modified version of the BSE simulator, which is *threaded*, i.e. uses multiple concurrently running computer processes ("threads") to model the parallel and

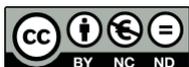

*1*

asynchronous nature of real financial markets, where multiple traders operate independently, in parallel and asynchronously: we call this simulator Threaded-BSE (TBSE). We then re-run our experiments on TBSE and find, as we show here for the first time, that the dominance hierarchy in TBSE is very different from that found in BSE: from our current results it seems that when operating in parallel the dominance hierarchy is instead ZIP>AA>ZIC>GDX. This is a novel result and is the key contribution of this paper. That is, we demonstrate here that previous use of simplistic simulation methodologies has produced sets of results which, although widely cited, can now no longer be trusted.

Section 2 of this paper gives a summary of the background to this work, and is taken largely verbatim from a position paper published by one of us at EMSS2019 (Cliff, 2019): our new results presented here can be read as an empirical illustration of the arguments made in that EMSS2019 paper. Section 3 gives more details of our methods; Section 4 shows key results; and Section 5 offers a discussion of the results, and of possible avenues of further work. TBSE is fully documented in (Rollins, 2020), including details of the TBSE source-code repository on GitHub.

## 2. Background

### 2.1. Traders, Markets, and Experimental Economics

The 2002 Nobel Prize in Economics was awarded to Vernon Smith, in recognition of Smith's work in establishing and thereafter growing the field of *Experimental Economics* (abbreviated hereafter to "EE"). Smith showed that the microeconomic behavior of human traders interacting within the rules of some specified market, known technically as an auction mechanism, could be studied empirically, under controlled and repeatable laboratory conditions, rather than in the noisy messy confusing circumstances of real-world markets. The minimal laboratory studies could act as useful proxies for studying real-world markets of any type, but one particular auction mechanism has received the majority of attention: the *Continuous Double Auction* (CDA), in which any buyer can announce a bid-price at any time and any seller can announce an offer-price at any time, and in which at any time any trader in the market can accept an offer or bid from a counterparty, and thereby engage in a transaction. The CDA is the basis of major financial markets worldwide, and tens of trillions of dollars flow through CDA markets every year.

Each trader in one of Smith's experimental CDA markets would be assigned a private valuation, a secret *limit price*: for a buyer this was the price above which he or she should not pay when purchasing an item; for a seller this was the price below which he or she should not sell an item. These limit-price assignments model the client orders executed by sales traders in real financial markets; we'll refer to them just as *assignments* in the rest of this paper. Traders in EE experiments from Smith's onwards are often motivated by payment of some form of real-world reward that is proportional to the amount of "profit" that they accrue from their transactions: the profit is the absolute value of the difference between the limit price specified when a unit is assigned to a trader, and the actual transaction price for that unit.

The limit prices in the assignments defined the market's supply and demand schedules, which are commonly illustrated in economics texts as supply and demand curves on a 2D graph with quantity on the horizontal axis and price on the vertical axis: where the two curves intersect is the market's theoretical competitive *equilibrium point:* indicating the *equilibrium price* (denoted here by $P_0$). A fundamental observation from microeconomics (the study of markets and prices) is that competition among buyers pushes prices up, and competition among sellers pushers prices down, and these two opposing influences on prices balance out at the competitive equilibrium point; a market in which transaction prices rapidly and stably settles to the $P_0$ value is often viewed by economists as *efficient* (for a specific definition of efficiency) whereas a market in which transactions consistently occur at off-equilibrium prices is usually thought of as inefficient: for instance, if transaction prices are consistently above $P_0$ then it's likely that buyers are being ripped off. By varying the prices in the traders' assignments in Smith's experiments, the nature of the market's supply and demand curves could be altered, and the effects of those variations on the speed and stability of the market's convergence toward an equilibrium point could be measured.

Smith's initial set of experiments were run in the late 1950's, and were described in his first paper on EE, published in the prestigious *Journal of Political Economy* (JPE) in 1962. The experiment methods laid out in that 1962 paper would subsequently come to dominate the methodology of researchers working to build adaptive autonomous automated trading agents by combining tools and techniques from Artificial Intelligence (AI) and Machine Learning (ML). This strand of AI/ML research converged toward a common aim: specifying an artificial agent, an autonomous adaptive trading strategy, that could automatically tune its behavior to different market environments, and that could reliably beat all other known automated trading strategies, thereby taking the crown of being the current best trading strategy known in the public domain, i.e., the "dominant strategy". Over the past 20 years the dominant strategy crown has passed from one algorithm to another and until very recently Vytelingum's (2006, 2008) "AA" strategy, was widely believed to be the dominant strategy, but recent results using contemporary large-scale computational simulation techniques indicate that it does not always perform so well as was previously believed, as discussed in the next section: Section 2.2 briefly reviews key publications leading to the development of AA, and the recent research that called its dominance into question.

### 2.2. A Brief History of Trading Agents

If our story starts with Smith's 1962 JPE paper, then the next major step came 30 years later, with a surprising result published in the JPE by Gode & Sunder (1993): this popularized a minimally simple automated trading algorithm now commonly referred to as *ZIC*. A few years later two closely related research papers were published independently and at roughly the same time, each written without knowledge of the other: the first was a Hewlett-Packard Labs technical report by Cliff (1997) describing the adaptive AI/ML trading-agent strategy known as the *ZIP* algorithm; the second summarized the PhD thesis work of Gjerstad, in a paper co-authored with his PhD advisor (Gjerstad & Dickhaut 1998), describing an adaptive trading algorithm now widely known simply as *GD*. After graduating his PhD, Gjerstad worked at IBM's TJ Watson Labs where he helped set up an EE laboratory that his IBM colleagues used in a study that generated world-wide media coverage when its results were published by Das *et al.* at the prestigious *International Joint Conference on AI* (IJCAI) in 2001. This paper presented results from studies exploring the behavior of human traders interacting with GD and ZIP robot traders, and demonstrated that both GD and ZIP reliably outperformed human traders. A follow-on 2001 paper by Tesauro & Das (two co-authors of the IBM IJCAI paper) described a more extensively *Modified GD* (MGD) strategy, and later Tesauro & Bredin (2002) described the *GD eXtended* (GDX) strategy. Both MGD and GDX were each claimed to be the strongest-known public-domain trading strategies at the times of their publication.

Subsequently, Vytelingum's 2006 thesis introduced the *Adaptive Aggressive* (AA) strategy which, in a major journal paper (Vytelingum *et al.*, 2008), and in later conference papers (De Luca & Cliff 2012a, 2012b), was shown to be dominant over ZIP, GDX, and human traders. Thus far then, ZIP had been beaten by GDX, and AA had beaten GDX, and hence AA held the title. In shorthand, we had AA>GDX>ZIP.

In all of the studies discussed thus far, typically two or three different types of trading algorithm would be compared against each other on the basis of how much profit (or *surplus*, to use the economists' technical term) they extract from the market, so Algorithm A was said to *dominate* or *outperform* or *beat* or *be stronger than* Algorithm B if, over some number of market sessions, traders running A made more money than traders running B. Methods of comparison varied. Sometimes a particular market set-up (i.e., a specific number of sellers, number of buyers, and their associated limit-price assignments specifying the market's supply and demand schedules) would be homogeneously populated with traders of type A, and then the same market would be re-run with all traders instead being type B, and an A/B comparison of profitability *in the absence of any other trading algorithms* could then be made. In other comparisons, for a market with *B* buyers and *S* sellers, B/2 of the buyers would use Algorithm A and the remaining B/2 buyers would run Algorithm B, with the seller population being similarly split, and the A/B comparison then showed profitability *in the presence of the other trading algorithm*. A/B tests involving 50:50 splits, as just described, were commonly used to establish the dominance relationship between A and B.

Comparatively recently, Vach (2015) presented results from experiments with the *OpEx* market simulator (De Luca, 2015), in which AA, GDX, and ZIP were set to compete against one another, and in which the dominance of AA was questioned: Vach's results indicate that whether AA dominates or not can be dependent on the ratio of AA:GDX:ZIP in the experiment: for some ratios, Vach found AA to dominate; for other ratios, it was GDX. Vach studied only a very small sample from the space of possible ratios, but his results prompted Cliff (2019) to use the public-domain "BSE" financial exchange simulator (BSE, 2012) to exhaustively run through a wide range of differing ratios of four trading strategies (AA, ZIC, ZIP, and the minimally simple SHVR built into BSE), doing a brute-force search for situations in which AA is outperformed by the other strategies. Cliff reported on results from over 3.4 million individual simulations of market sessions, which indicated that Vach's observation was correct: whether AA dominates does indeed depend on how many other AA traders are in the market, and what mix of what other strategies are also present. Depending on the ratio, AA could be outperformed by ZIP and by SHVR. Subsequent research by Snashall (2019) employed the same exhaustive testing method, using a supercomputer to run more than one million market simulations (all in BSE) to exhaustively test AA against IBM's GDX strategy: this again revealed that AA does not always dominate GDX: see Snashall & Cliff (2019) for discussion.

In this paper we will talk about counting the number of "wins" when comparing an A algorithm to a B algorithm: in the experiments reported in Section 4, we create a specific market set-up, and then run some number *n* of independent and identically distributed markets sessions with some ratio A:B of the two strategies in the buyers, and the same A:B ratio in the sellers. In any one of those sessions, if the average profit per trader (APPT) of type A traders is higher than the APPT for traders of type B, then we count that session as a "win" for A; and *vice versa* as a win for B. In the experiments reported in Section 4, assignments to buy or sell are issued to the traders periodically, with all traders being updated at the same time, and the limit prices in the assignments came from symmetric supply and demand curves where the equilibrium price was varied dynamically using the sinusoidal offset function illustrated in Fig.4.8 of the BSE user guide (BSE, 2012).

Our experiments reported here are motivated by this progression of past research. In particular, we noted that Vach's results which first revealed that the ratio of different trading algorithms could affect the dominance hierarchy came from experiments he ran using De Luca's (2015) OpEx market simulator, which is a true parallel asynchronous distributed system: OpEx involves a number of individual trader computers (discrete laptop PCs) communicating over a local-area network with a central exchange-server (a desktop PC). But many of the other results that we have just summarized came from financial-market simulators that only very roughly approximated parallel execution: Cliff (1997) published C-language source-code for the discrete-event simulator he developed to test and compare ZIC with ZIP; and the BSE simulator (BSE, 2012) also uses a very simple time-sliced approach where, if any one trader is called upon to issue a response to a change in the market, it always does so within exactly one simulated time-slice within the simulation, regardless of how much computation it has to execute to generate that response. Snashall & Cliff (2019) noted that the actual reaction-times of the various trading algorithms varied quite widely, and in a true parallel simulation the slower traders might be expected to do much less well than when they are evaluated or compared in a temporally simplistic simulation. So, possibly Vach's results were as much to do with his use of OpEx as his varying of the ratios of trader-algorithms. That is what we set out to explore in this paper. In order to do that, we developed TBSE, a new threaded (parallel) version of the BSE financial-market simulator, discussed in the next section.

## 3. TBSE: Threaded BSE

Much of the functionality of TBSE was directly borrowed from the original BSE simulator. The key changes affect the way each market session operates. In BSE, each market session is executed on a single thread in which a core loop executes repeatedly until the session end-time is passed. Within this loop a single trader in the market's population of traders is selected at random, and that trader is polled to see if it wants to issue an order: this is achieved by calling that trader's `getorder` function. If the trader does issue an order, that order is then processed by the exchange and all traders in the population are notified of any resultant change in the market-data published by the exchange. At that point, all traders in the population are given an opportunity to update their internal values with the new information, via a function called `respond` that is particular to each specific trading algorithm (that is, `respond` is where much, perhaps all, of the detail of the trading algorithm is encoded); and then control loops back to another random selection of a single trader to be polled for an order. A single pass through this sequence of steps is implied to take a notional time of $1/N$ seconds, where *N* is the total number of traders in the market: in this way, each trader will on average get a chance to issue an order once per simulated second, and will call respond *N* times per simulated second. But if Trader A's `respond` takes 1ms of real-time to execute, and trader B's `respond` takes 10s of real-time to compute an answer, BSE's record of simulated time is the same in both cases: the simulated clock advances in increments of $1/N$ seconds, regardless of the wall-time consumed by the differing `respond` functions.

The problem with this style of simulation is that each trader is allotted as much time as it needs to execute its `getorder` and `respond` functions, and it is guaranteed that nothing else in the market will change while it executes either of these functions. This means that the execution time of each trader has no impact on its performance: instead all that counts is its ability to generate an order price that will be accepted by another trader, whilst attempting to generate the greatest profit. This differs from a real financial exchange where all traders operating on the market will be operating asynchronously. In an asynchronous market, a trader may look at the exchange's market data, i.e. the currently available information on that market, and use this to calculate what it considers its best order to send to the exchange. However, if this trader uses a complex, slow running algorithm to do so, it may find that by the time it has completed its calculation, another simpler and faster trader has already succeeded in executing a trade, which as a result changes the position of the market which may render the order than the complex trader has posted less profitable than expected. This situation, which happens all the time in real-world financial markets, cannot be modelled in BSE.

To address this, we created an asynchronous threaded exchange simulator, TBSE. BSE was written in the *Python* programming language, and so we used that for TBSE too. In TBSE each trader executes on its own computational thread, as does the exchange. There is also the main thread of the Python code which continues to run during the market session and is responsible for the distribution of customer orders to each of the traders. Each trader thread consists of a loop which executes continuously until the end of the trading session. Within this loop the trader first receives information of trades which have been successfully executed at the exchange, it then updates its own internal record keeping if any of these trades involved itself and executes its `respond` function to update its internal variables based on the new market data. Once it is updated with the latest market data it then executes its `getorder` function which determines whether it should post a new order to the exchange, and if so at what price. This order is then put on a queue to be sent to the exchange.

The exchange operates on this queue, reading orders from the queue and then processing them by either adding them to the its list of available orders or executing a trade if the new order can be matched with any of its current list of available orders. If a trade can be executed it then places the details of the resultant transaction onto a queue for each trader to read before submitting its next order. The system for processing orders and trades is almost identical to that of BSE.

In Python, as with other languages that offer such a feature, *multi-threading* is where a processor can operate multiple threads of execution concurrently. This differs from multi-processing, which Python also supports, where different processes execute simultaneously. Both are forms of parallelism, but whereas in multi-processor programming multiple processes are executing at the same time on different physical processors, in multithreading only one thread can execute at any one time. The execution of each thread is divided into short segments and the processor regularly switches between each thread for a short period of time, giving the appearance that all threads are executing at the same time. It is not guaranteed that each thread will be given the exact same amount of execution time during each cycle, but over the course of executing an entire program, which may involve millions of cycles, the execution time given to each thread should tend towards the same value. In TBSE this means that each trader is given the same amount of processing time, so a faster algorithm can complete its execution and start processing a second order while a slower algorithm is still calculating its first order price. In Python, the execution of threads is synchronized by a Global Interpreter Lock (GIL) which ensures that only one thread is executing at a time. Multi-threading was chosen over multi-processing as most general-purpose CPUs have no more than 8 processing cores, but for our experiments we want to simulate the parallel execution of many more than 8 traders operating simultaneously.

The queues used within TBSE are synchronous first in, first out (FIFO) queues, meaning that the trader who places an order on the exchange queue first will have their order processed first. This is a good simulation of real exchange behavior as it is commonplace for exchanges to prioritize orders by their time of arrival.

## 4. Results

Table 1 shows a high-level summary of our results for AAvsZIC, AAvsZIP, GDXvsZIC, and GDXvsZIP. Characterizing each of these four experiments as Algorithm A vs Algorithm B, the central sub-table of Table 1 shows the count of "wins" for A and the count of wins for B in BSE, with the higher of the two counts highlighted in bold font; and then the right-hand sub-table of Table 1 shows the corresponding win-counts for A and for B operating instead in TBSE, again with the higher value highlighted in bold font.

As can be seen from Table 1, our results for BSE are consistent with those previously published in the literature: there is nothing in our BSE results to challenge the status quo: AA>ZIC & ZIP; GDX>ZIC&ZIP. However, the TBSE results in Table 1 tell a very different story. AA still beats ZIC, which is intuitively what one would expect. But the other three dominance relationships have been inverted: in TBSE, ZIC beats GDX while ZIP beats both AA and GDX – scoring more than twice as many wins as GDX, beating it by a larger margin than it beat ZIP in BSE.

If we chose, we could stop here. We have now established an answer to the question that we set out to explore: whether a switch to a more realistic (i.e., parallel) market simulation makes a difference to the dominance relationships previously reported: it clearly does. This is a significant finding and is the primary contribution of this paper. However, there is more to say.

Each number in Table 1 represents the number of "wins" scored by a specific algorithm, in a series of 19x500 market sessions, ranging over 19 ratios of TraderA:TraderB (denoted by $R_a$: ranging here from 1:19 to 19:1), with n=500 trials at each value of $R_a$. One obvious thing to do is to separate the aggregate results for any one pair of trader algorithms into the 19 sets of results, one for each value of $R_a$, and to look for any interesting changes in the patterns of wins as $R_a$ is swept from one extreme to another. Table 2 shows such a data-set, for AA-vs-ZIC. As you can see, at the bottom of the table is the sum of all the wins in each column, and the relevant column-sums in Table 2 are the values that populate the first row of Table 1. Table 2 adds a third column to the results for BSE and TBSE: a variable we call *Δwins*, which is simply the difference between the two algorithm's win-counts. In an A-vs-B comparison of two trading algorithms: if $\Delta wins>0$ then A outperforms B; and if $\Delta wins<0$ then B outperforms A. The two sets of $\Delta wins$ values shown in Table 2 can be plotted graphically in the style shown in Figure 1, a paired plot that we refer to as the *delta curves* for two trading strategies tested in BSE and then in TBSE.

**Table 1.** Summary of all our experiment results, showing total number of "wins" summed over 19 different A-vs-B trials in BSE (central sub-table) and in TBSE (right-hand sub-table). The 19 different trials vary the ratio $R_a$ of trading algorithm A ("AlgoA") to trading algorithm B ("AlgoB") from 1:19 through 10:10 to 19:1, and at each ratio we conduct n=500 i.i.d market sessions, which are treated as contest between AlgoA and AlgoB: if, at the end of a session, the average profit per trader for AlgoA is greater than that for AlgoB, then that counts as a "win" for AlgoA. Hence, the maximum possible score is 19x500=9,500. In each row, bold-font text is used to highlight the larger number of wins in each A-vs-B comparison. As can be seen, switching from BSE to TBSE has no effect in the case of AA-vs-ZIC, but for the other three cases we see a reversal of the dominance relationship. See text for further discussion.

| AlgoA | AlgoB | BSE # A Wins | BSE # B Wins | TBSE # A Wins | TBSE # B Wins |
|---|---|---|---|---|---|
| AA | ZIC | **7095** | 2405 | **7370** | 2130 |
| AA | ZIP | **7581** | 1739 | 4383 | **5117** |
| GDX | ZIC | **5517** | 3988 | 4300 | **5199** |
| GDX | ZIP | **6538** | 2962 | 2574 | **6926** |

The rest of this results section shows selected highlights from digging deeper into the outcomes of our experiments. As it happens, a deeper dig unearths some thought-provoking results.

The delta curves from Table 2, i.e. for AA-vs-ZIC, look roughly the same as each other, modulo some noise from the stochastic elements in our simulation, and are not shown here, but Figures 1, 2, and 3 show the delta curves for AA-vs-ZIP, GDX-vs-ZIC, and GDX-vs-ZIP, respectively. These are the three sets of experiments in which the switch from BSE to TBSE inverted the dominance relationship between the trading algorithms, and they deserve some examination and discussion.

**Table 2.** The data that was aggregated into the top row of Table 1, tabulated to show the individual results from each of the 19 different ratios of the two algorithms used: n=500 at each ratio. The full table is not shown here, because the specific details do not matter. Column sums are displayed at the bottom of the table, and correspond to the values given in the top row of Table 1.

| Ratio | BSE AA Wins | BSE ZIC Wins | BSE $\Delta wins$ | TBSE AA Wins | TBSE ZIC Wins | TBSE $\Delta wins$ |
|---|---|---|---|---|---|---|
| 1:19 | **279** | 221 | 58 | **297** | 203 | 94 |
| 2:18 | **355** | 145 | 210 | **357** | 143 | 214 |
| 3:17 | | | 204 | **375** | 125 | 250 |
| | | | | **384** | 116 | |
| 17:3 | **373** | 127 | | | | |
| 18:2 | **361** | 139 | 222 | | | 324 |
| 19:1 | **324** | 176 | 148 | **346** | 154 | 192 |
| Sum | **7095** | 2405 | 4690 | **7370** | 2130 | 5240 |

**AA-vs-ZIP** (Figure 1). Here there seems to be some coherent structure in the BSE delta curve: although AA consistently outperforms ZIP, the degree by which it beats ZIP seems to attenuate when the $R_a$ is at either extreme of its range, and the maximum outperformance of AA over ZIP seems to be when $R_a<10:10$, i.e. when AA is in the minority of the population. The relevance of this is seen in the TBSE delta curve in Fig.2: we saw in Table 1 that ZIP wins on aggregate in this set of experiments, but Fig.2 makes clear that when AA is in a small minority in TBSE it can still outperform ZIP, yet as soon as the ratio of AA traders exceeds 50% its dominance disappears and it is outperformed by ZIP.

**GDX-vs-ZIC** (Figure 2). Here, for both BSE and TBSE, the delta curves have a clear coherent structure to them, which is something that we do not think has been reported before in the literature: we know from Vach (2015), Cliff (2019), and Snashall & Cliff (2019) that ratio matters, but none of those publications reported or explored such strongly coherent relationships between ratio and results, between $R_a$ and $\Delta wins$.

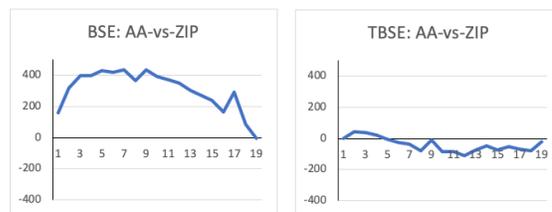

**Figure 1.** "Delta Curves" for AA vs ZIP in BSE (left) and TBSE (right). This pair of graphs shows the $\Delta wins$ data, as was tabulated for AA vs ZIC in Table 2: the horizontal axis shows $R_a$ and is labelled with the number of AA traders in the ratio; the vertical axis is the value of $\Delta wins$. When $\Delta wins>0$, AA outperforms ZIP, and when $\Delta wins<0$, ZIP beats AA.

Figure 2 shows fairly unambiguously that GDX outperforms ZIC when GDX is in the minority, and ZIC outperforms GDX when ZIC is in the minority. Figure 3 also reveals that the qualitative nature of the swing from GDX dominance in BSE to ZIC dominance in TBSE differs from that from AA to ZIP that was illustrated in Figure 1: whereas the two delta plots in Figure 1 are markedly different, the two curves in Figure 2 are remarkably similar: the TBSE curve could plausibly be described as what happens when the BSE curve is shifted slightly to the right and slightly down. Also notable is that in both BSE and TBSE the maximum outperformance of ZIC by GDX happens at roughly a ratio of 1:3, and the maximum outperformance of GDX by ZICs seems similarly to happen at roughly 3:1. This strikes us as curious in that experience tells us that usually in these kind of experiments maxima would normally be expected to occur either at the endpoints of the scale (i.e. at ratios of 1:19 or 19:1) or near the midpoint (i.e. at 10:10) why this is so is something we aim to investigate in further work.

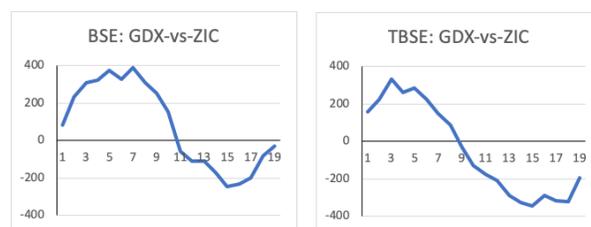

**Figure 2.** Delta curves for GDX vs ZIC; format is the same as in Figure 1.

**GDX-vs-ZIP** (Figure 3): again, there is a clear coherence to the delta curves, although the relationship between the BSE and TBSE curves is not as similar as that in Figure 2, and not as different as that in Figure 1. In a point of similarity with the GDX-vs-ZIC curves, again the peak performance of the two strategies come at ratios of roughly 3:1 and 1:3 (and, again, we do not know why this should be so) but for GDX-vs-ZIP the relative performance of the two algorithms does not level out to zero near ratio values of 10:10 and then cross into underperformance for the algorithm that is in the majority; instead in BSE GDX pretty consistently outperforms ZIP and in TBSE the situation is the inverse: now ZIP is the dominant algorithm at almost all ratios, and again the degree of dominance falls steadily as the proportion of ZIPs increases beyond 1:3). Further work is required to understand why these delta curves have this particular qualitative shape.

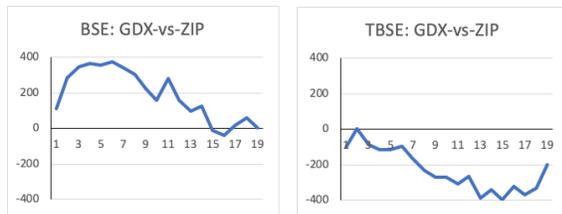

**Figure 3.** Delta curves for GDX vs ZIP; format is the same as in Figure 1.

## 5. Discussion and Further Work

One thing that is notable from the delta curves in Figures 1, 2, and 3 is that, despite the points of similarity highlighted above, the effect that shifting from BSE to TBSE has on the delta curves is qualitatively different for each pair of trading algorithms: the data we have studied thus far reveals no conveniently simple patterns that allow us to make a priori predictions or generalizations about when Algorithm A will outperform Algorithm B. Figure 2 is the starkest illustration of this point: recall that GDX was developed at IBM TJ Watson Research Labs, was preceded by two earlier versions (GD, then MGD), and when published was described by IBM as the best-performing trading algorithm in the then-published literature. We do not seek to criticize the IBM team, but our results show that, when the conditions are right (i.e., when the ratio of GDX:ZIC is in the right range), in fact GDX can be consistently out-performed by ZIC. Gode & Sunder's ZIC paper had been published six years before IBM's GDX paper, and while GDX mixes the construction of a probabilistic belief function with techniques from dynamic programming, the ZIC algorithm takes up one line of code.

One avenue of further work comes in attempting to understand what features, if any, of the trading algorithms interact in such a way that they give rise to the delta curves that we have plotted here, and the extent to which those delta curves are affected by changes in other significant factors, such as the market's supply and demand schedules. Also, here we presented results that focus on four trading algorithms that have been used repeatedly in studies of artificial trading systems over the past 20 years, but there are several other algorithms which could be added into this analysis. And, when we're done with analyzing interactions between A/B pairs of trading algorithms, we can move on to A/B/C triples (in which case the delta curves could be plotted as points on a simplex), and then on to various ratios of four or five or six different algorithms, etc. But as the number of algorithms involved in any one comparison increases, so do the number of trials required (the combinatorics are explosive, and the computational cost even of the experiments shown here was measured in days of CPU time), and so do the difficulties of visualizing and analyzing the results.

## 6. Conclusion

Prior to this paper, anyone reading the trading-agent literature would have been likely to form the opinion that there was widespread agreement that, in general, AA beats GDX, GDX beats ZIP, and ZIP beats ZIC. Only a careful reading of the literature would reveal that many of the relevant results came from single-threaded simulations. Our results presented here show that while the AA>GDX>ZIP>ZIC dominance hierarchy may hold true in simple simulations, as soon as real-time factors matter, i.e. as soon as the various algorithms are operating in parallel, the computational costs of a sophisticated algorithm such as GDX count against it, and it can be outrun by simpler but faster algorithms. So, the primary contribution of this paper is our demonstration here that the old single-threaded dominance hierarchy is not maintained in multi-threaded TBSE. As Table 1 shows, in A/B comparisons on TBSE we have AA beating ZIC, ZIP beating AA, ZIC beating GDX, and ZIP beating GDX; results that can be summarized as ZIP>AA>ZIC>GDX.

Our work calls into question not just the truth of specific claims of dominance, but also whether it is ever worth trying to make such claims at all, because any trading-algorithm's performance, and hence dominance, is clearly so heavily affected by factors exogenous to that trading algorithm, chief of which is what other algorithms it is competing against, and in what proportion or ratio those different algorithms are present in the market: it's a manifestly game-theoretic situation, but game theory offers no help here. We question whether, in markets that are sufficiently realistic to be relevant to the real world, there can ever really be a single specific trading algorithm that is "dominant", that actually beats all the rest. Hence, we conclude with this: if you think your trading algorithm really is the dominant one, you have probably tested it in simulations that are too simple.